\documentstyle[12pt]{article}
\begin{document}
\newcommand{\mycheck}[1]{\marginpar{#1}}
\newcommand{\ot}{\frac{1}{2}}
\newcommand{\D}{& \displaystyle}  
\newcommand{\di}{\displaystyle}   
\newcommand{\SH}{S\!H}
\font \math=msbm10 scaled \magstep 1
\newcommand{\mmath}[1]{{\mbox{\math #1}}}
\renewcommand{\Re}{\mmath{R}}
\newcommand{\ZZ}{\bf \math Z}

\date{July 1994}
\title
{Compact U(1) Gauge Theory on Lattices with Trivial
Homotopy Group}

\author{C. B. Lang\\ \\
{\normalsize Institut f\"{u}r Theoretische Physik}\\
{\normalsize Universit\"{a}t Graz}\\
{\normalsize A-8010 Graz, Austria}\\ \\
and\\ \\
T. Neuhaus\\ \\
{\normalsize Fakult\"at f\"ur Physik}\\
{\normalsize Universit\"{a}t Bielefeld}\\
{\normalsize D-33615 Bielefeld, Germany} }

\maketitle
\begin{abstract}

We study the pure gauge model on a lattice manifold with trivial
fundamental homotopy group, homotopically equivalent to an $S_4$.
Monopole loops may fluctuate freely on that lattice without
restrictions due to the boundary conditions. For the original Wilson
action on the hypertorus there is an established two-state signal in
energy distribution functions which disappears for the new geometry.
Our finite size scaling analysis suggests stringent upper bounds on
possible discontinuities in the plaquette action. However, no
consistent asymptotic finite size scaling behaviour is observed.

\end{abstract}
\thispagestyle{empty}
\newpage

\section{Motivation}

Pure gauge lattice QED is a prototype model of gauge interactions of
abelian nature and was among the first gauge systems studied in Monte
Carlo simulations \cite{CrJaRe79,LaNa80,DeTo80}.  It is also a starting
points of the discussion of the more realistic lattice field theory
containing dynamical fermion degrees of freedom. Unfortunately the
status of nonperturbative studies of full QED is still controversial.
In this situation it is important to clarify the fixed point properties
i.e., the critical properties and critical exponents of the pure gauge
system.

{}From the early beginnings this field theoretic lattice model has
posed many interesting questions about its continuum limit. The compact
formulation of the pure gauge model has a confining strong coupling
phase and a spin-wave (Coulombic) phase. The order of the phase
transition has been controversial throughout the years. However, it was
generally accepted that the formulation with the Wilson action on a
finite lattice leads to a clear two-state signal at the transition,
indicating a first order deconfinement phase transition
\cite{JeNeZe83}.

Adding extra terms to the action introduces new couplings. In this
larger space the nature of the phase transition may change. Allowing
for an adjoint coupling $\gamma$ (multiplying the plaquette variable in
the adjoint representation \cite{BhCr81}) a clear increase of the
energy gap towards positive $\gamma$ was observed \cite{EvJeNe85}.
Based on a detailed study of discontinuities of the plaquette
expectation value in the $\beta -\gamma$ plane and, using a power law
scaling ansatz for the vanishing of discontinuities along the first
order phase transition line, the location of a possible change of the
order of the deconfinement phase transition to a second order phase
transition line was estimated \cite{JeNeZe83,EvJeNe85}.  From that
extrapolation the infinite volume location of the so-called tricritical
point in pure gauge compact QED was expected to lie at slightly
negative values of the coupling $\gamma$. Thus the theory on the Wilson
line was thought to have a weak first order phase transition, however
close to a tricritical point in the extended parameter space.

Let us stress here, that up to now {\em no} direct simulation of pure
compact QED did unanimously show the existence of a second order phase
transition point. It is well known from statistical mechanics systems
that sizeable crossover effects from a discontinuous to a continuous
behaviour are present on finite lattices in the vicinity of a
tricritical point. Such effects are also present in pure gauge compact
QED even at reasonably large negative values of $\gamma$ and they spoil
attempts to extract critical exponents of the system using finite size
scaling methods.  However, due to the central role of the model as a
starting point of full QED and due to its prototype character in
lattice gauge studies, it would be very important to know e. g. the
value of the correlation length divergence exponent $\nu$.  A possible
determination then could tell us whether $D=4$ pure gauge compact QED
might be in the universality class of the free theory with $\nu=0.5$,
or not. We may remark here that MCRG studies of the model showed in the
vicinity of the Wilson line a flow of couplings, which in spite of the
two state signal resembled the flow diagram of a second order critical
point quite well \cite{La86}.

The nature of the deconfinement phase transition appears to be
inherently connected to topological excitations present in this
formulation (where the gauge fields assume their values in the group
$U(1)$, not in the algebra).  Formulated on the dual lattice these are
closed loops of monopoles.  In several investigations a strong
correlation between the behaviour of the monopole loop density with the
internal energy variable was observed
\cite{DeTo80,GrJaJe85,BaSh,BhLiSc,BoMiMu}.
If the monopole loops are forbidden or suppressed by
further terms in the action \cite{BaSh,BoMiMu} a shifting
of the phase transition towards smaller values of the plaquette
coupling $\beta$ was found.

Here we concentrate on a modification of the simulation, which appears
to be softer than introducing additional terms to the action, and which
should become irrelevant in the thermodynamic limit. Instead of
choosing periodic (hypertorus) boundary conditions we work on a lattice
topology homotopic to the surface $S^4$ of a 5D hypersphere. In the
infinite volume limit the two versions should lead to equivalent
results.

There is a subtle difference, which, however, may be important. The
monopole loops are nonlocal objects. The Monte Carlo update algorithms
are intrinsically local (except for scarce attempts towards multigrid
updating). This is related to the gauge symmetry nature of the field
variables. On the hypertorus closed loops lie in equivalence classes
defined by the fundamental homotopy group of the torus: $\Pi_1(T^4)=
\math Z^4$. This puts some restrictions on the fluctuation behaviour of
such loops. Hypercooling a hot configuration shows that monopole loops
wrapped around the torus may got stuck like rubberbands
\cite{GrJaJe85}. Recently this behaviour was studied and classified in
more detail \cite{KeReWe94}.  On the other hand, the fundamental
homotopy group of the surface of a sphere is (for $D>2$) trivial:
$\Pi_1(S^4)= I$. There monopole loops may contract and disappear
without restriction.

Preliminary results with half the statistics and for $\gamma=0$ have
been published earlier\cite{LaNe94a}.

\section{Action and Lattice Geometry}

The action on the lattice $\Lambda$ is given
\begin{equation}
S = \beta \sum_{P\in \Lambda} \mbox{Re} (U_P) + \gamma
\sum_{P\in \Lambda}
\mbox{Re} (U_P^2)
\end{equation}
where $U_P$ denotes the usual plaquette variable and we consider all
plaquettes with equal weight (see, however the discussion at the end).

For the construction we choose the $4$-manifold as the boundary of a 5D
hypercube $H_5(L)$ with linear extend $L$ ($L^5$ sites). We denote this
boundary by $\SH_4(L)$; it consists of all sites which have at least
one cartesian coordinate value $1$ or $L$. The manifold $\SH_4(L)$ has
\begin{equation}\begin{array}{lcl}
\mbox{sites}  &:& 10(L-1)^4+20(L-1)^2+2  \\
\mbox{links}  &:& 40(L-1)^4+40(L-1)^2  \\
\mbox{plaquettes} &:& 60(L-1)^4+20(L-1)^2\\
\mbox{3-cubes} &:& 40(L-1)^4\\
\mbox{4-cubes} &:& 10(L-1)^4
\end{array}
\label{geo}
\end{equation}
thus ratios of e.g. $n_{\mbox{\footnotesize
links}}/n_{\mbox{\footnotesize sites}}$ for the $\SH_4(L)$ lattice
approach for large $L$ asymptotically the values for the hypercubic
geometry.  Whereas the hypertorus lattice $T_4(L)$ is self-dual,
$\SH_4(L)$ is obviously not.  For an identification of the monopole
loops we did construct the dual manifold $\SH'_4(L)$ (which has the
same homotopy properties) as usual, identifying the dual sites with the
centers of the hypercubes of the original lattice. This was done to
check the dynamics of the updating algorithm.

We define the volume of the lattice $V\equiv|\Lambda|$ as the {\it
number of plaquettes} and we use $V^{\frac{1}{4}}$ as the linear size
variable; as finite size scaling is asymptotic this definition is as
arbitrary as any other.  In the computer programs the geometrical
properties have been implemented by tables providing vectorization
possibilities.

\section{Simulation and Coupling Parameters}

We performed long runs with an integrated statistics of 500K up to 1500K
sweeps for each lattice size, using a combination of 3-hit Metropolis
updating with one overrelaxation step.

Lattices $\SH(L)$ were studied for $L=4,6,8,10$ having roughly the same
number of variables as $6^4\ldots 16^4$ lattices with torus geometry.
Most of our data was taken for values of $\gamma=0$ and $\beta$ close
to the transition.

Simulations were done for various coupling values close to the
pseudocritical values.  For given lattice size the histograms from
different couplings were combined to one multi-histogram according to
the technique of Ferrenberg and Swendsen \cite{FeSw}.
Here we introduce statistical mechanics notations and define the
partition function by
\begin{equation}
Z_L(\beta) = \sum_E \rho_L(E)\exp{(-\beta E)},
\end{equation}
where we denote the internal energy function by $E = -\sum_P{\rm
Re}(U_P)$. Note that $E$ is extensive and negative for positive
$\beta$-values. Correspondingly we can introduce energy densities
$e=E/V$, which up to a negative sign correspond to the plaquette
operator.  Here $\rho(E)$ denotes the density of states for each
value of $E$ and is determined from the combined individual
distributions. In our simulation we determine expectation values of the
specific heat and two cumulants \cite{ChLaBi86,Bi92}
\begin{eqnarray}\label{cumdefs}
C_V(\beta,L)&=& ~~\frac{1}{V} \langle (E-\langle E\rangle )^2
\rangle ,
\nonumber \\
V_{BCL}(\beta,L)&=& -\frac{1}{3} \frac{\langle (E^2-\langle
E\rangle^2)^2\rangle}{\langle E^2\rangle^2} ,\\
U_4(\beta,L)&=& \frac{\langle (E-\langle E\rangle)^4\rangle}
{\langle(E-\langle E\rangle )^2\rangle^2} \nonumber .
\end{eqnarray}
Our definition of $V_{BCL}$ differs by an additive constant
$\frac{2}{3}$ from that of \cite{ChLaBi86}.  The specific heat $C_V$
develops a maximum, while $V_{BCL}$  and $U_4$ both develop minima as
functions of $\beta$ for given lattice size. We define pseudocritical
couplings through the peak positions of these quantities and peak
values of the considered quantities by their values at
pseudocriticality.

We observed quite long autocorrelation lengths at the corresponding
pseudocritical points for each lattice size, with peak values of
$\tau_{int}$ ranging from 15 (for $L=4$) up to 750 (for $L=10$).  This
was one of the reasons for doubling the statistics and introducing
overrelaxations steps as compared to the preliminary results presented
in \cite{LaNe94a}.

\section{Results and scaling analysis}
\subsection{Pseudocritical couplings}

Table 1 gives the pseudocritical couplings and the values of the
cumulants at the peak positions.

In fig. 1 we compare the multihistograms resulting from all data for
given $L$, that is $h_L(\beta,e=E/V)\equiv\rho_L(E)\exp{(-\beta E)}$ at
the values of the peak positions of $C_V$. These are the effective
probability distributions for the action at these couplings. We find
{\it no} indication of double peaks, in distinct contrast to results
for the same action but standard hypertorus geometry
\cite{JeNeZe83,BhLiSc,LaNe94a,AzDiGr90}.

We then performed runs at $\gamma=0.2$ where torus-results had
identified even more prominent two-state signals. An analogous
multihistogram analysis there showed pronounced two-state signals for
the theory on $\SH$ lattices even on the smallest lattice sizes.
Whenever this situation occurs, tunneling becomes scarce on larger
lattices and the multihistogram analysis becomes impracticable. In this
case one should rely on multicanonical simulations \cite{BeNe}. For
that reason we did not attempt simulations for $\SH(L>6)$ at this value
of $\gamma$.

Returning to $\gamma=0$ we display in fig. 2 as a function of the
variable $1/V$ a comparison of pseudocritical $\beta$-values on $\SH$
lattices, namely $\beta(C_V), \beta(V_{BCL})$ and $\beta(U_4)$, with
results from simulations on the standard hypertorus geometry for
$\beta(C_V)$ \cite{Ne94x}.  We find a strikingly different finite size
scaling behaviour when comparing both sets of data. The pseudocritical
$\beta$-values of the $\SH$ lattices approach the infinite volume
critical $\beta_c$-value from above, while data on the hypertorus do so
from below. Furthermore our data indicate in case of $\beta(V_{BCL})$ a
non-monotonic behaviour on the $\SH$ lattice, while on the standard
hypertorus the approach seems to be monotonic for all quantities.

In principle finite size scaling theory gives the asymptotic approach
of the considered pseudocritical $\beta$-values.  If the deconfinement
phase transition is a continuous phase transition the homogeneous free
energy density $f_{hom}(\beta)=f_{hom}({{\xi(\beta)\over L}})$ scales
with the scaling variable; with the usual definition of the correlation
length divergence exponent $\nu$ one has for large
volume systems the asymptotic behaviour
\begin{equation}
\beta_{c,V^{1\over 4}}  =\beta_c + a V^{-{1\over{4\nu}}}.
\label{critical}
\end{equation}
Alternatively, if the phase transition is of discontinuous nature (first
order), we expect an expansion of the pseudocritical $\beta$-values in
powers of the inverse volume $1/V$ \cite{BoKo,LeKo91}
\begin{equation}
\beta_{c,V}  =\beta_c + {a\over V} + {\cal O}(1/V^2).
\label{discont}
\end{equation}
In fig. 2 we display a fit to the $\beta(C_V)$-values on the standard
hypertorus according to the scaling law (\ref{discont}) including
$1/V^2$ corrections (dashed curve in the figure).  This fit gives a
consistent description of the data and results in the infinite volume
$\beta_c$-estimate $\beta_c=1.01102(9)$. In the current context this
demonstrates  the consistency of the data with a finite size scaling
behaviour of a discontinuous transition in the considered volume size
interval.  As already mentioned, this was anticipated for the case of
the hypertorus geometry. That fit alone is not sufficient, though.  A
definite statement on whether the deconfinement phase transition is of
first order type would require the calculation of its latent heat and
interface tension in the thermodynamic limit.  Such an analysis is
beyond the scope of our present work. Later on we argue, indeed, that
evidence for a weaker first order or even continuous phase transition
can be obtained from $\SH$ lattices.

Returning to the theory on $\SH$ lattices we note that by construction
the theory is {\it not} homogeneous. On the surface of the 5D hypercube
there are inhomogeneities related to its edges i.e., the theory on the
$\SH$ lattice exhibits less symmetries than formulated on a
hypertorus.  In general one then expects additional finite size scaling
violations which may be parametrized by a contribution
$f_{inhom}(\beta)$ to the free energy density on the $\SH$ lattice.  In
the thermodynamic limit $f_{inhom}(\beta)$ will vanish.

In absence of a rigorous theory we may estimate the dimensionality of
the additional terms by considering quantities sensitive to the
inhomogeneity of the lattice. For instance we might consider ratios of
the number of plaquettes to the number of links (\ref{geo}), or the
number of links which are contained in $4$ or $5$ different plaquette
variables as compared to the standard case where each link is contained
in $6$ different plaquette variables.  These considerations suggest a
dimensionality of the inhomogeneous free energy density part $\propto
1/V^{\ot}$.

The usual ansatz for the partition function (the indices $o$ and $d$
denote ordered and disordered phase) of a discontinuous
system, neglecting interfacial effects, is
\begin{equation}
Z(\beta) = \exp{(-V f_d)} + A \exp{(-V f_o)}.
\end{equation}
Here the quantity $A$ parametrizes the asymmetry. The
free energy densities $f_o(\beta)$ and $f_d(\beta)$ can be expanded
in powers of $\Delta\beta =\beta-\beta_c$
\begin{eqnarray}
f_o&=&f_{o,hom}+f_{o,inhom}\nonumber\\
&=&c+e_o \Delta \beta + V^{-\ot}
(\eta_o + \sigma_o \Delta \beta) + {\cal O}((\Delta \beta)^2)
\end{eqnarray}
(and correspondingly for $f_d$).  In the infinite volume limit
$f_o=f_d=c$ at $\Delta \beta =0$. At finite volume we find at the
metastability point, where both contributions have same weight,
\begin{eqnarray}
{\rm ln} A &+& e_o \Delta \beta + V^{-\ot}
(\eta_o + \sigma_o \Delta \beta) =\nonumber\\
&&e_d \Delta \beta + V^{-\ot}
(\eta_d + \sigma_d \Delta \beta)+ {\cal O}((\Delta \beta)^2)\;.
\end{eqnarray}
Solving for $\Delta \beta$ we then expect
the finite size scaling law of (\ref{discont}) to be replaced by the form
\begin{equation}
\beta_{c,V}  =\beta_c + {a\over V^{\ot}}
+{b\over V}+{\cal O}(V^{-{3\over2}}),
\label{inhomog}
\end{equation}
if the phase transition was of discontinuous nature.
It turns out that the dominant term now is $O(1/V^{\ot})$.

In fig. 2 we show fits of the pseudocritical $\beta$-values for the
$\SH$ lattices according to (\ref{inhomog}) (dotted curves).  These
fits result in a positive contribution proportional to $1/V^{\ot}$ and
a negative contribution proportional to $1/V$, which actually is of the
same order of magnitude as in case of the hypertorus. Thus the observed
non-monotonic behaviour observed in the quantity $\beta(V_{BCL})$ could
be explained by the interplay of a contribution related to the
inhomogeneity, entering with the opposite sign, as compared to a
contribution $\propto 1/V$, which may be attributed to the homogeneous
part of the free energy density.  The obtained infinite volume
$\beta_c$-estimates from these fits $\beta_c(CV)=1.01133(13)$ ,
$\beta_c(V_{BCL})=1.01134(15)$ and $\beta_c(U4)=1.01125(16)$ are
consistent with the result from the hypertorus, though a tendency
towards values slightly higher by about one standard deviation is
noticeable.  Again we emphasize that this analysis is merely a
statement on the consistency of different finite size scaling fits in
the considered and limited volume size interval i.e., both
regularisations on the hypertorus and the $\SH$ lattice support a
unique infinite volume $\beta_c$-value.  As far as the $\SH$ lattice is
concerned, we conclude, that the behaviour of the pseudocritical
$\beta$-values is strongly affected by contributions stemming from the
inhomogeneity of the system. Consequently we have to pay special
attention to the effect of these contributions when in the later
analysis we study finite volume estimators of energy gaps at the
deconfinement phase transition.

We have also analyzed the $\SH$ lattice pseudocritical $\beta$-values
assuming the finite size scaling form (\ref{critical}) corresponding to
a critical point. Including data with $L \geq 8$ into the fit we obtain
$\nu$-values of about $\nu \approx 0.53(20)$ and $\beta_c$-values
consistent with the above determinations.  However, as discussed such a
value of $\nu$ close to $0.5$ can just be mimicked by contributions
related to the inhomogeneity of the system (\ref{inhomog}). We
therefore cannot trust these fits, as long as these contributions are
sizable and likely to be present. This observation also indicates that
an unambiguous extraction of critical exponents of the deconfinement
phase transition, if it were in fact a continuous transition, would
require much larger lattice sizes, or alternatively, a formulation of
the theory with smaller finite size scaling violations.

\subsection{Cumulant peak values}

The peak values of the specific heat and cumulants defined in
(\ref{cumdefs}) give in the thermodynamic limit unambiguous signals on
the order of the considered phase transition.  If the phase transition
is of discontinuous nature one finds in the thermodynamic limit ${L\to
\infty}$ on general grounds
\begin{eqnarray}
{C_{V,max}\over V}&=&  ~~{(e_o-e_d)^2\over 4} \; ,\nonumber \\
V_{{BCL},min}  &
=& -{(e_o^2-e_d^2)^2\over 12(e_oe_d)^2} \; ,\\
U_{4,min}  &=& 1\; . \nonumber
\end{eqnarray}
Here $e_o$ and $e_d$ denote the discontinuous values of the energy
density at the critical point. In the context of the present paper
$-e_o$ corresponds to the plaquette expectation value in the Coulomb
phase, $-e_d$ to its expectation value in the confinement phase. For a
continuous phase transition $e_o=e_d$. In this case $U_{4,min}$ can be
different from $1$.

Again finite size scaling theory predicts for discontinuous and {\em
homogeneous systems}, that finite size corrections to above quantities
on finite volumes follow asymptotically an expansion in powers of the
inverse volume $1/V$, similar to the expansion written down in
(\ref{discont}).  In fig. 3a we display a comparison of results on the
hypertorus (triangles) with results on $\SH$ lattices (circles) for the
quantity $C_{V,max}(L)/ V$ as a function of the parameter $1/V$. This
quantity can be directly interpreted as a finite volume estimator of a
possible infinite volume discontinuity in the internal energy or the
plaquette expectation value, correspondingly.  We observe for both
geometries a deviation from the leading asymptotic finite size scaling
with terms proportional $1/V$ i.e., higher order terms in the
corresponding expansions are important.  It is therefore rather
doubtful, whether a fit to the data, even including higher orders,
could lead to a unambiguous determination of the infinite volume
discontinuities.

Here we are not that ambitious.  We note, however, that the $\SH$ data
exhibit much lower values for the finite volume estimator of the
discontinuity on comparable volume sizes. Thus, if the observed finite
size scaling on the $\SH$ lattice stays monotonic in the variable $1/V$
up to to the thermodynamic limit, then the $\SH$ results predict rather
stringent $\it upper~bounds$ on the values of possible discontinuities
at the phase transition point. In light of the above discussion on
possible effects related to the inhomogeneous part of the free energy
density we have to be rather careful.  We therefore performed in our
analysis several fits to the observable ${C_{V,max}(L)/V}$ including
$1/V^{\ot}$, $1/V$ and also corrections proportional to $1/V^{3 \over
2}$. In none of these cases we found indication of a non-monotonic
behaviour. The dotted curve of fig. 3a displays as an example such a
fit with the form
\begin{equation}
{C_{V,max}\over V} ={(e_o-e_d)^2\over 4} + {a\over V^{\ot}} + {b\over V}.
\end{equation}
Thus we convert the value obtained on the $SH(10)$-lattice,
${C_{V,max}(10)/ V}=.0000528(13)$, to an upper bound on the possible
discontinuity in $e_d-e_o$ in the thermodynamic limit
\begin{equation}
e_d-e_o \leq 0.0145(2) ,
\label{bound}
\end{equation}
which then should be satisfied by pure gauge compact QED with the
Wilson action in the hypertorus geometry as well as on $\SH$ lattices.
We remark here that such a bound is more stringent than any gap
observed or extrapolated from torus results \cite{EvJeNe85};
simulations on the hypertorus on $16^4$ lattices typically result in
estimates of the discontinuity of about $0.030$ on that lattice size.
In a study of the discontinuity in the $\beta - \gamma$ plane of
couplings a power law ansatz suggested an infinite volume value of the
above discontinuity of about $0.016$ \cite{EvJeNe85}.  However the
statement here is that (\ref{bound}) constitutes a bound and that the
true value of the discontinuity is likely to be still smaller.  Thus
compact pure gauge QED on $\SH$ lattices provides evidence that the
first order deconfinement phase transition on the Wilson line is at
least weaker than previously thought.

In fig. 3b we also display a comparison of results on the hypertorus
(triangles) with results on $\SH$ lattices (circles) for the quantity
$V_{{BCL},min}$. The remarks concerning deviations from asymptotic
finite size scaling in the preceding paragraph are appropriate here as
well. Performing various fits of the already mentioned forms
(dotted curve in fig. 3b) we do not find a non-monotonic behaviour.
Again the observed monotonic behaviour of the quantity $V_{{BCL},min}$
lead to an upper bound on possible discontinuities.  On the $SH(10)$
lattice (cf. table 1) we obtain the inequality
\begin{equation}
e_d-e_o \leq -0.022(1) \, e_d.
\label{relation}
\end{equation}
The plaquette expectation value in the disordered state $-e_d$ can be
easily bounded from above; numerical simulation results give $-e_d \leq
0.66$. Inserting this number we again obtain $e_d-e_o \leq 0.0145(2)$,
which is precisely the bound derived in (\ref{bound}).  Finally we
display in fig. 3c the quantity $U_4$. It does not exhibit asymptotic
finite size scaling behaviour for either version of the theory and it
is not possible to decide whether it approaches a value of $1$ (for a
first order phase transition) or larger (continuous transition).

All together our findings indicate that asymptotic finite size scaling
for both formulations, on the hypertorus and on the $\SH$ lattice, is
very far away from the lattice sizes, which we are able to investigate
in our present studies.  This is not surprising for the theory on the
hypertorus as we know from studies in much simpler models (like Potts
models) with first order phase transitions, that asymptotic behaviour
may set in only on very large lattices \cite{BiNeBe93} and in general
little is known about the onset of finite size scaling for gauge
theories. On $\SH$ lattices additional finite size corrections are
caused by the inhomogeneity.

\section{Discussion}

The studied geometry -- the boundary of a 5D hypercube -- has less
symmetry than the hypertorus. Asymptotically it amounts essentially to
a sum over a set of ten hypercubes (like in 2D it corresponds to the
six faces of the cube, the boundary contributions become negligible).
On finite lattices the reduced translational and lattice rotational
symmetry gives rise to large additional contributions of finite size
scaling violating operators.  Thus the asymptotic scaling behaviour
sets in only on very large lattice sizes. A possible cure for the
observed finite size scaling violations may be the introduction of
weight factors i.e., considering the metric for a truly curved topology
like the $S_4$. Studies in this direction including also terms
proportional to $\gamma$ into the action are under progress and we
expect from those studies clarification on the critical properties of
pure gauge compact QED close to the Wilson line.

It remains a remarkable fact that the clear two-state signal of earlier
Monte Carlo simulations disappears when one changes the lattice
geometry with respect to their properties at the boundary. This finding
corresponds to a rather small bound on possible discontinuities in the
plaquette operator, which, based on the observed monotonic behaviour of
finite volume discontinuities, is suggested by the present analysis.
It also remains a remarkable fact, that the introduced soft
modification of the theory leads in comparison to the results for the
theory on a hypertorus to such large effects.  This may be related to
the influence of closed loops for the dynamics of the phase transition
and may be a special property of gauge theories containing the compact
$U(1)$ gauge group. One may wonder whether this observation can be of
relevance for other systems with loop excitations, including systems
with fermionic fields.

\bigskip

{\bf Acknowledgement:}

In the final stage of our work we profited from discussions with Jiri
Jers\'ak. C.B.L. wants to thank Martin L\"uscher for a discussion.
T.N. is indebted to Thomas Lippert and Klaus Schilling for a useful
discussion on the subject.

\newpage
{\noindent \Large \bf Tables}

\vspace{0.5cm}
{\noindent \bf Table 1:~}
Peak positions and values of the measured cumulants.
\smallskip
\begin{center}
\begin{tabular}{llll}
\hline
$L$     &       $C_{V,max}$     &$V_{BCL,min}
\times 10^3$&$U_{4,min}$\\
\hline
4       &       2.21(1)         &-1.631(7)     &2.739(8)\\
6       &       4.96(13)        &-0.45(11)     &2.625(30)\\
8       &       10.08(32)       &-0.23(8)      &2.455(43)\\
10      &       20.91(62)       &-0.17(5)      &2.263(37)\\
\hline
 \\
\hline
$L$   & $\beta (C_V)$   &       $\beta (V_{BCL})
$       &$\beta (U_4)$\\
\hline
4   &   1.01536(22)     &       1.01194(29)   & 1.01688(37)\\
6   &   1.01370(16)     &       1.01326(18)   & 1.01401(19)\\
8   &   1.01263(6)      &       1.01257(6)   &  1.01273(7)\\
10   &  1.01216(1)      &       1.01215(1)   &  1.01218(2)\\
\hline
\end{tabular}
\end{center}

\newpage
\begin{figure}[htb]
\begin{center}
\hfil \hbox to 12cm{\vbox to 14cm{\vfil
 } }\hfil
\end{center}

\caption{
The multihistograms $h_L(\beta(C_V),e)$ resulting from all data for
given $L$ at the values of the pseudocritical points as defined by the
peak positions of the specific heat.  Shown are data on
$\SH(4),\SH(6),\SH(8)$ and $\SH(10)$ lattices.
}
\end{figure}

\newpage
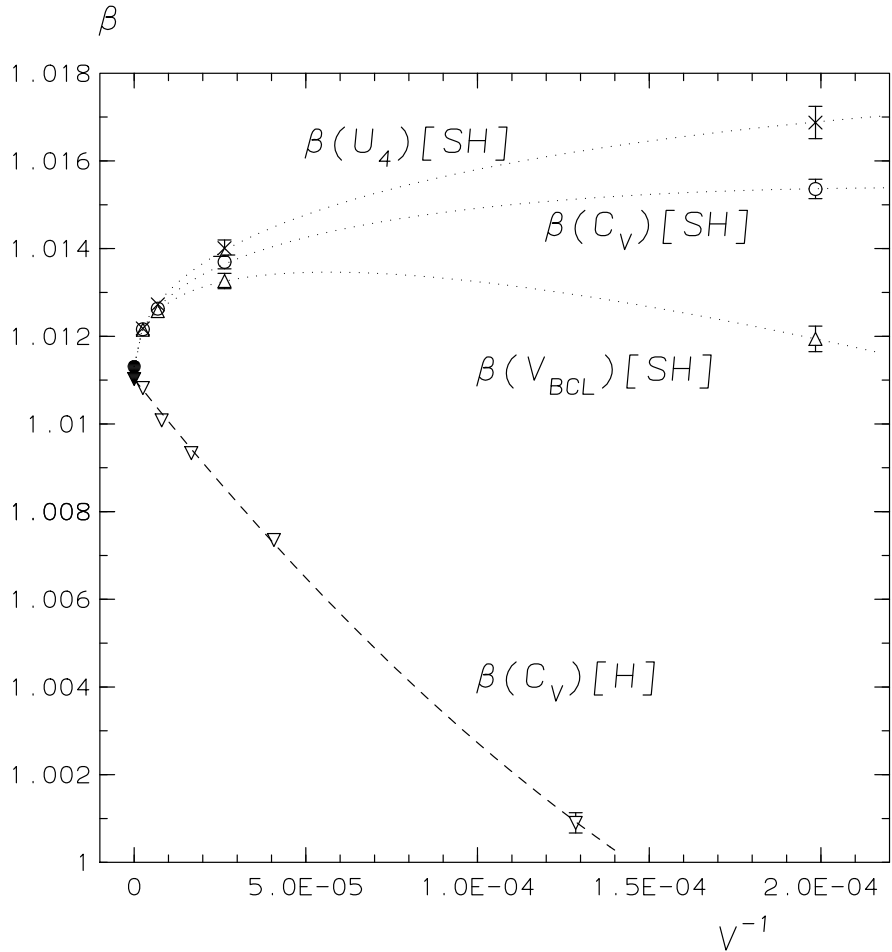
\begin{figure}[htb]
\begin{center}
\hfil \hbox to 12.5cm{\vbox to 12cm{\vfil
 } }\hfil
\end{center}
\caption{
The pseudocritical $\beta$-values as a function of the parameter $1/V$.
We compare results for $\beta(CV)$ (circles), $\beta(V_{BCL})$
(triangles) and $\beta(U4)$ (crosses) on the $\SH$ lattice with results
from simulations for $\beta(CV)$ (triangles down) on the hypertorus.
The data on the hypertorus have been obtained in multicanonical
simulations \protect{\cite{Ne94x}} on lattices ranging in volume
between $6^4$ and $12^4$.  On the hypertorus we include also a data
point from \protect{\cite{BhLiSc}} on the $16^4$ lattice.
The dashed curve represents a fit according to
(\protect{\ref{discont}}), the dotted curves represent fits according
to (\protect{\ref{inhomog}}).  The full symbols represent infinite
volume extrapolations of $\beta_c$.
}
\end{figure}

\newpage
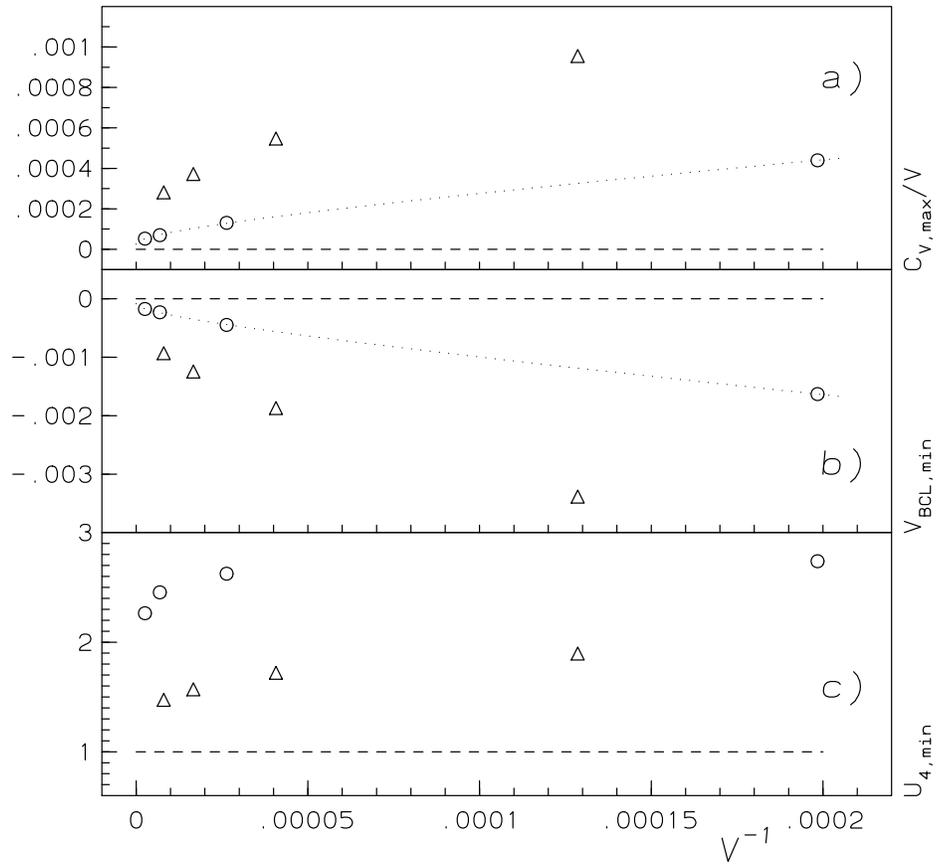
\begin{figure}[htb]
\begin{center}
\hfil \hbox to 12.5cm{\vbox to 14cm{\vfil
 } }\hfil
\end{center}
\caption{
The peak values of (a) specific heat $C_V$  and of (b) the cumulants
$V_{BCL}$  and (c) $U_4$ vs. $1/V$.  We compare data on the hypertorus
(triangles) with data on $\SH$ lattices (circles). The dotted curves
are explained in the text.
}
\end{figure}

\end{document}